\documentclass[%
 reprint,showkeys,
 amsmath,amssymb,
 aps,
pra,
]{revtex4-1}

\usepackage{graphicx}% Include figure files
\usepackage{dcolumn}% Align table columns on decimal point
\usepackage{bm}% bold math

\usepackage{amsmath,amssymb,amsfonts}%
\usepackage{amsthm,physics}%
\usepackage{mathrsfs}%
\usepackage{tikz}
\usetikzlibrary{quantikz}
\usepackage{comment}
\usetikzlibrary{positioning}
\usetikzlibrary{arrows.meta}
\usetikzlibrary{shapes.arrows}
\newtheorem{theorem}{Theorem}
\newcommand\independent{\protect\mathpalette{\protect\independenT}{\perp}}
\def\independenT#1#2{\mathrel{\rlap{$#1#2$}\mkern2mu{#1#2}}}
\newtheorem{proposition}[theorem]{Proposition}
\newtheorem{definition}[theorem]{Definition}

\newtheorem{assumption}[theorem]{Assumption}
\begin{document}
%\preprint{APS/123-QED}

\title{Quantum interpretations, causality and quantum computation}% 
\author{Vivek Kumar}
\email{vivekk.phd20.cs@nitp.ac.in}
\affiliation{%
 CSE, NIT Patna, PU Campus\\ Patna, 800005, Bihar, India}
\author{M P Singh}%
\email{mps@nitp.ac.in}
\affiliation{%
 CSE, NIT Patna, PU Campus\\ Patna, 800005, Bihar, India}
\author{R Srikanth}
\email{srik@ppisr.res.in}
\affiliation{
 Theoretical Sciences, PPISR \\
 Bidalur post, Devanahalli, Bengaluru 562164, Karnataka, India
}%

%\date{\today}% It is always \today, today,
             %  but any date may be explicitly specified

\begin{abstract} 
The interpretation of quantum mechanics continues to be debated, and quantum nonlocality accentuates the puzzle. Quantum interpretations can be classified broadly into two types: \textit{realist interpretations}, which assert that quantum states describe objective reality (even if hidden or branching), and \textit{subjective interpretations}, which treat quantum states as observer-dependent information or beliefs about the system. Here we study the implication of quantum interpretations for causal explanations of Bell nonlocal correlations, and show that a given interpretation type carries an inherent commitment to a preferred causal structure.  Specifically, we find that realist interpretations entail a classical causal model, and thus require Fine-Tuning to prevent superluminal signaling, while subjective interpretations are found to entail a framework of nonclassical causal models. The implications of our results for one-way quantum computation and computation-based Bell nonlocality are studied.
\end{abstract}

%Use showkeys class option if keyword
                              %display desired
\maketitle
\section{\label{sec:intro} Introduction}
Despite quantum theory's success in passing several stringent experimental tests and in applications in modern technology, there is a lack of consensus among physicists and philosophers on how the formalism of quantum mechanics may be interpreted to represent reality \cite{Cabello_2017, leifer2014quantum}. The debate about the nature of the quantum state centers broadly about two positions: (1)  \textit{Realist interpretations} which assert that quantum states describe objective reality (even if hidden or branching). Among realist interpretations are various objective collapse models \cite{bassi2021philosophy, ghirardi1986unified, diosi1989models, penrose1996gravity, patel2017weak}, pilot-wave mechanics \cite{goldstein2001bohmian}, the Many-Worlds interpretation \cite{vaidman2002many, dewitt2015many}, as well as the $\psi$-epistemic model of Spekkens \cite{spekkens2007evidence};  (2) \textit{Subjective interpretations}, which treat quantum states as observer-dependent knowledge or beliefs of the system.  Among subjective interpretations, we may count the more common version of standard Copenhagen \cite{faye2002copenhagen}, QBist \cite{fuchs2014introduction}, the no-interpretation thesis \cite{fuchs2000quantum} and information-based \cite{zeilinger1999foundational, brukner2017quantum} interpretations. Our above bifurcation corresponds roughly to Cabello's classification of type I and II of interpretations \cite{Cabello_2017}.

On an entirely different line of inquiry, a modern approach to causal analysis has been through the framework of causal models \cite{pearl2009causality}, which extends 
Reichenbach’s principle of common cause \cite{reichenbach1991direction} to a more general causal structure, represented by a directed acyclic graph (DAG). This principle asserts that the correlation between two observed variables should have a classical causal explanation in terms of a direct influence from one variable to the other or a common cause that conditionally factorizes the joint probability of the variables in the sense of classical probability theory. The framework of causal models generalizes the principle to the case of a joint probability distribution among multiple variables as constrained by causal relationships among them. It is usually acknowledged that classical causal models  are inadequate for explaining quantum correlations unless we include fine-tuned mechanisms such as superdeterminism, retrocausality or superluminal influences \cite{wood2015lesson, cavalcanti2018classical}. As a consequence, a major research effort has recently been into the investigation of nonclassical models of causal inference. These models generalize the Reichenbach common cause principle \cite{cavalcanti2014modifications, cavalcanti2021implications, allen2017quantum, coiteux2021no, lauand2023witnessing}, while averting the Fine-Tuning problem.

Moving from the original conceptual critique of Fine-Tuning \cite{wood2015lesson} to a computationally verifiable proof of incompatibility,
Navascu\'es \& Wolfe \cite{navascues2020inflation} 
%provide a general, algorithmic advance over the original, conceptual arguments of Wood \& Spekkens \cite{wood2015lesson} by 
introduced the inflation technique, which uses convex optimization to rigorously test the compatibility of any data with a classical causal model.
%moving from a philosophical critique of fine-tuning to a . 
Ref. \cite{barrett2021cyclic} goes beyond the usual acyclic quantum causal models to study cyclic quantum causal models, which tackle causality in timelike correlations with feedback loops.
%in contrast to the usual acyclic quantum causal model works concerned with explaining spacelike correlations (such as Bell nonlocality) via quantum common causes. 
Starting with \cite{bong2020strong}, various works have studied \textit{local friendliness} (LF) in a quantum causal setting, arguing that LF violations imply a need to extend relativity to events themselves- making events relational rather than absolute \cite{cavalcanti2021implications, wiseman2023thoughtful, ormrod2024quantum}. More recently, Liu et al. \cite{liu2025quantum} present a protocol for quantum causal inference identifying causal relationships and temporal order between two quantum systems using only minimal, non-invasive (coarse-grained) quantum measurements.%, without requiring active interventions or reset operations.

%Ref. \cite{ormrod2024quantum} introduces \textit{event relativity}, a framework within quantum causal models that defines quantum events (such as measurement outcomes) as relative to observers' causal contexts using projectors, resolving paradoxes like Wigner’s friend and explaining decoherence-like effects without relying on absolute quantum states. This approach advances quantum foundations by providing a consistent, relational description of quantum phenomena tied to causal structures.} 

Prima facie, it would appear that any framework of causal models addresses only aspects of the quantum formalism, and is oblivious to the question of the interpretation of quantum mechanics. However, here we will find that a given quantum interpretation can have a preferred framework. The wider implications of this result are studied by means of its application to Bell nonlocality in the context of quantum computation.
%Yet there is some basis to expect that the theories or models in these frameworks may impact or be influenced by certain aspects of the interpretation problem. For example, quantum cognition shows that quantum-like structures can appear in human decision making \cite{pothos2022quantum}. In real life, there can a subjective element to causality in that our intuition and relative perception of a factor can influence whether we infer that it caused some event \cite{quillien2020we}.

The remaining sections are structured as follows: In Section \ref{sec:causmod}, we briefly review classical and causal models of quantum correlations applicable to the Bell scenario. In Section \ref{sec:reality}, we present our main results pointing out that realist and subjective interpretations have distinct implications for the causal structure of quantum correlations. The ramification of these results for certain quantum computational situations is discussed in Section \ref{sec:1wqc}, where we consider the causal structure of one-way (or, measurement-based) quantum computation, as well as of a Bell test on certain many-particle states that are classically hard to compute.  Finally, we present our conclusions and discussions in Section \ref{sec:discus}. 

\section{\label{sec:causmod} Classical and quantum causal models and quantum correlations}
Bell inequalities can be derived from a conjunction of classical and relativistic causality \cite{bell1976theory, wood2015lesson}. This provides an alternative derivation to the traditional one based on the assumption of local-realism \cite{bell1964on}. Therefore, a violation of a Bell inequality entails that either classical causality or relativistic causality must be abandoned. Below we make precise these two concepts in the context of the two-input-two-output-two-party Bell scenario. One of the principle assumptions underlying classical causality is the Reichenbach principle of common cause \cite{reichenbach1991direction}, which can be broken down into the following two assumptions.
\begin{definition}
[\bf Principle of common cause (CC)] 
If two events $A, B$ are non-trivially correlated, i.e., if $P(A, B) \ne P(A)P(B)$, then either:
\begin{enumerate}
    \item There is a direct causal connection between them, i.e. either $A$ causes $B$ or vice versa.
    \item A and B share a common cause in their causal past that fully accounts for the correlation.
\end{enumerate}
\end{definition}
The following definitions clarify the concepts of a common cause and the causal past.

\begin{definition}\textnormal{\bf (Factorization of probabilities (FP)).} Two events $A$ and $B$ have a common cause ($\Lambda$) if and only if $\Lambda$ corresponds to events in the two events' common causal past such that conditioning on $\Lambda$ decorrelates the two events:
\begin{equation}
    P(A,B|\Lambda) = P(A|\Lambda)P(B|\Lambda),
\label{eq:fp}
\end{equation}
or a mixture of such factorized probabilities. 
\label{def:fp}
\end{definition}
The causal past that appears in the characterization of the PCC framework depends on the spacetime causal structure assumed.
The conjunction of CC and FP is the Reichenbach principle of common cause. The Causal Markov Condition of classical causal modeling \cite{pearl2009causality} formalizes the idea of the Reichenbach common cause influencing multiple variables in a Bayesian network, asserting that a variable is independent of its non-descendants conditioned on its parents (i.e., direct causes) in DAGS (Appendix \ref{sec:DAG}).

For our immediate purpose, it suffices to characterize classical causality in the context of Bell scenario as follows:
\begin{definition}[\bf Classical causality] Because the Reichenbach principle is encompassed by classical causality, we have:
\begin{equation}
    \textnormal{Classical}~ \textnormal{causality} \Longrightarrow \textnormal{CC} + \textnormal{FP}.
    \label{eq:classicalc}
\end{equation}
\label{def:classicalc}
\end{definition}
Relativity theory requires identifying the causal past with the past light cone.
\begin{definition}[\textbf{Relativistic causality (RC)}]
   All events happen within a single relativistic space-time, in which all causes of a given event $A$ will be found in the past light cone of $A$.
   \label{def:rcs}
\end{definition}
The conjunction of classical and relativistic causalities is Bell's concept of \textit{local causality} \cite{bell1976theory}. In other words,
\begin{equation}
\text{CC} + \text{FP} + \text{RC} \Longrightarrow \text{Bell's}~ \text{inequalities}
\label{eq:rp+rcs}
\end{equation}
The causal structure appropriate to the Bell scenario subject to the constraint imposed by local causality is depicted in Fig. \ref{fig:bell}. Here, circular nodes represent random variables corresponding to observed events, such as measurement inputs or outputs, while square nodes represent a latent resource. Directed edges indicate direct causes.
\color{black}

By the causal Markov condition of classical causality (Appendix \ref{sec:DAG}), any distribution $P(AB|XY)$ that is compatible with the DAG Fig. \ref{fig:bell} must satisfy the conditional independences of $P(B|Y,X)=P(B|Y)$ and $P(A|X,Y)=P(A|X)$, which represent the no-signaling conditions. Conversely, by the principles of causal discovery, starting from the above conditional independences together with the independence of the inputs, namely $P(X,Y)=P(X)P(Y)$, we can infer the DAG Fig. \ref{fig:bell}, provided the principle of \textit{No Fine-Tuning} is assumed. This principle requires that the causal weights of the microscopic causal influences shouldn't be carefully to ensure that the signal vanishes at the macroscopic level. 

\begin{figure}
    \centering
    \begin{tikzpicture}[scale=1.5]
  % Nodes
  \node[draw, shape=rectangle, minimum width=1cm, minimum height=1cm, fill=gray!20, font=\large] (lambda) at (0, 0) {$\mathbf{\Lambda}$};
  \node[draw, shape=circle, minimum size=1cm, font=\large] (A) at (-1.5, -1.5) {\textbf{A}};
  \node[draw, shape=circle, minimum size=1cm, font=\large] (B) at (1.5, -1.5) {\textbf{B}};
  \node[draw, shape=circle, minimum size=1cm, font=\large] (X) at (-1.5, -3) {\textbf{X}};
  \node[draw, shape=circle, minimum size=1cm, font=\large] (Y) at (1.5, -3) {\textbf{Y}};

  % Arrows
  \draw [-{To[length=2mm, width=3mm, fill=none]}] (lambda) -- (A);
  \draw [-{To[length=2mm, width=3mm, fill=none]}] (lambda) -- (B);
  \draw [-{To[length=2mm, width=3mm, fill=none]}] (X) -- (A);
  \draw [-{To[length=2mm, width=3mm, fill=none]}] (Y) -- (B);
  
  %\draw[->] (lambda) -- (A);
  %\draw[->] (lambda) -- (B);
  %\draw[->] (X) -- (A);
  %\draw[->] (Y) -- (B);
\end{tikzpicture}
\caption{\textbf{DAG for the Bell scenario motivated by relativistic causality.} Here the square node represents a nonclassical common cause, with the shading to indicate that it represents a quantum entangled state or a generalized probabilistic correlation.}
\label{fig:bell}
\end{figure}

Quantitatively, the LHS in Eq. (\ref{eq:rp+rcs}) entails that
\begin{equation}
    P(A,B|X,Y) = \sum_{\Lambda=\lambda} p(\lambda)P(A|X,\lambda)P(B|Y,\lambda),
\label{eq:localcausality}
\end{equation}
where $\sum_{\lambda} p(\lambda)=1$. Such a distribution necessarily satisfies the Clauser-Horne-Shimony-Holt (CHSH) version of Bell's inequality \cite{clauser1969proposed}:
\begin{align}
    P(A=B|0,0) &+ P(A=B|0,1) + P(A=B|1,0) \nonumber \\
     &+ P(A\ne B|1,1) \le 3.
     \label{eq:BI}
\end{align}
A number of experiments have shown that the inequality (\ref{eq:BI}) is violated. From Eqs. (\ref{eq:classicalc}) and (\ref{eq:rp+rcs}), we find that an explanation in the framework of classical causality would require superluminal causes (Figure \ref{fig:Arrow}), and thus require rejecting RC. In turn, this would necessitate unwelcome Fine-Tuning in order to hide the superluminal causal influence on the macroscopic statistical level \cite{wood2015lesson, cavalcanti2018classical}.
%i.e., the no-signaling is an epistemic effect of our ignorance of the underlying superluminal causal influences. 

In response to this impasse, various frameworks of quantum causality provide a way out, preferring to abandon the classical causality (specifically, FP), rather than relativistic causality. For our purpose, this will be the defining characteristic of any framework of quantum causality.
\begin{definition}[\bf Quantum causality] Abandoning FP and hence Reichenbach principle, but adhering to CC, we have:
\begin{equation}
    \textnormal{Quantum}~ \textnormal{causality} \Longrightarrow \textnormal{CC} + \textnormal{RC}.
\label{def:quantumc}
\end{equation}
\end{definition}
For the Bell scenario, this position entails a DAG consistent with relativistic causality, as in Fig. \ref{fig:bell}, except that $\Lambda$ can be something more general than shared classical randomness.

Broadly, quantum causality modifies the assumption FP in two ways. In the interventionist approach, such as in Ref. \cite{allen2017quantum}, the common cause doesn't require the factorization of probabilities, as in Eq. (\ref{eq:fp}), but instead of Choi states, obtained via the Choi-Jamiolkowski quantum channel-state isomorphism. In the theory-independent approach, the common cause is not restricted to a classical resource but can be a shared quantum resource, such as a Bell state or (in the context of a generalized probability theory \cite{barrett2007information}) a PR box. Table \ref{tab:classical_vs_quantum} summarizes the essential distinguishing features of classical vs quantum causality as applied to the Bell scenario.

\begin{table}[]
    \centering
    \begin{tabular}{|c|c|c|c|c|}
    \hline
            & CC & FP & RC & Fine- \\
            &    &    &    & Tuning \\
             \hline
       Classical causality  & yes & yes & no & yes \\
       \hline
        Quantum causality   & yes & no  & yes & no\footnote{In the sense of absence of multiple mutually canceling strategies for causal explanation \cite{srikanth2025}.} \\
        \hline
    \end{tabular}
    \caption{Classical vs quantum causality applied to the violation of Bell's inequality.}
    \label{tab:classical_vs_quantum}
\end{table}

\begin{figure}
    \centering
   \begin{tikzpicture}[scale=1.5]
  % Nodes
  \node[draw, shape=rectangle,, minimum width=1cm, minimum height=1cm, font=\large] (lambda) at (0, 0) {$\mathbf{\Lambda}$};
  \node[draw, shape=circle, minimum size=1cm, font=\large] (A) at (-1.5, -1.5) {\textbf{A}};
  \node[draw, shape=circle, minimum size=1cm, font=\large] (B) at (1.5, -1.5) {\textbf{B}};
  \node[draw, shape=circle, minimum size=1cm, font=\large] (X) at (-1.5, -3) {\textbf{X}};
  \node[draw, shape=circle, minimum size=1cm, font=\large] (Y) at (1.5, -3) {\textbf{Y}};

  % Arrows
  \draw [-{To[length=2mm, width=3mm, fill=none]}] (lambda) -- (A);
  \draw [-{To[length=2mm, width=3mm, fill=none]}] (lambda) -- (B);
  \draw [-{To[length=2mm, width=3mm, fill=none]}] (X) -- (A);
  \draw [-{To[length=2mm, width=3mm, fill=none]}] (Y) -- (B);
  \draw [-{To[length=2mm, width=3mm, fill=none]},double] (X) -- (B);
  \draw [-{To[length=2mm, width=3mm, fill=none]}, double] (A) -- (B);
\end{tikzpicture}
    \caption{\textbf{DAG for the Bell scenario motivated by classical causality.} Here the square node represents a classical common cause. The violation of a Bell inequality can be explained by a superluminal signal from Alice's input to Bob's output, represented by the double-edge. Fine-Tuning of the edges $X\rightarrow B$ and $X \rightarrow A \rightarrow B$ is required to enforce no-signaling in the observed probabilities.}
    \label{fig:Arrow}
\end{figure}

\section{\label{sec:reality} Causal structure and quantum interpretations}
In general, making the notion of a realist or subjective interpretation mathematically precise is difficult. Certain kinds of interpretation can be mathematically defined and conclusions drawn about them within a limited framework, as for example the question of reality vis-\`a-vis the $\psi$-epistemic vs. $\psi$-ontic interpretations in the Harrigan-Spekkens ontological framework \cite{harrigan2010einstein}. For our purpose, the following will suffice.

\begin{definition}[\bf Realist interpretation.] The position that the quantum state corresponds to an objective property of the physical world, and reflects an underlying physical reality that exists independent of observers or measurements. Measurement outcome probabilities relect epistemic ignorance about an underlying reality or intrinsic stochasicity built into Nature.
\label{def:realist}
\end{definition}

\begin{definition}[\bf Subjective interpretation] 
A subjective interpretation takes the quantum state as a description of an observer's knowledge, beliefs, or information about a system — not a direct reflection of objective physical reality.
\label{def:subjective}
\end{definition}
The following result is essentially a retelling of Bell's theorem through a causal lens. The key point is that each interpretation commits to what principle underlying Bell's concept of local causality should be abandoned in order to explain Bell nonlocality.

%\textcolor{red}{FOR EASY REFERENCE1. Main Results (Propositions 8 and 9 in Section III): These propositions are relatively simple consequences of Definitions 6 and 7 and are already well understood by researchers in the field. Specifically, it is well known that violations of Bell inequalities prevent the unsteerable assemblage description given in Eq. (7); hence, Alice’s measurement necessarily affects Bob’s assemblage. In the proof of Proposition 8, the authors state: “By Definition 6, the quantum state is real, and hence so is the disturbance of Bob’s state due to Alice’s measurement X.” This implies that there must be a causal influence from X to B, which in turn requires fine-tuning, as discussed, for example, in Ref. [18].In the proof of Proposition 9, the authors write: “By Definition 7, the quantum state represents the observer’s knowledge or belief about the system’s state. Therefore, the remote measurement disturbance produced by Alice constitutes an epistemic rather than real causal influence.” As a result, relativistic causality can be preserved, but at the cost of abandoning classical notions of causality. These observations are conceptually clear and follow naturally from the assumptions, limiting their novelty.}

\begin{proposition}
[Quantum interpretation and causal structure] In any given realist (resp., subjective) interpretation of QM, a causal explanation for Bell nonlocality requires a classical (resp., quantum) causal structure.
\label{thm:onto}
\end{proposition}
\begin{proof}
Let Alice and Bob verify a Bell inequality violation while performing measurements in a common inertial frame of reference. The distribution $P(AB|XY)$ produced by Alice's initial and Bob's subsequent measurements on a joint state $\rho_{\mathcal{AB}} \equiv \sum_{\Lambda=\lambda} p(\lambda) \rho_{\mathcal{AB}}^{\lambda}$ is in general given by:
\begin{equation}
    P(A,B|X,Y) = \sum_{\Lambda=\lambda} p(\lambda)P(A|X,\lambda)\textrm{Tr}_{\mathcal B}(\Pi_{B|Y}\rho_{\mathcal B}^{(A| X,\lambda)}),
\label{eq:steerable}
\end{equation}
where $\sum_\lambda p(\lambda)=1$, $\Pi_{B|Y}$ is the projector for measuring $Y$ and obtaining outcome $B$, and $\rho_\mathcal{B}^{(A|X,\lambda)}$ is Bob's conditional state. 

Consider a subjective interpretation of Eq. (\ref{eq:steerable}). Upon Alice performing a projective measurement $X$ on her local Hilbert space $\mathcal{H}_{\mathcal A}$ with outcome $A$, she leaves the joint system in the post-measurement state $\rho_{\mathcal{AB}}^{(A|X,\lambda)} \equiv \frac{\Pi_{(A|X)} \rho_{AB}^{\lambda} \Pi_{(A|X)}^\dagger}{P(A|X,\lambda)}$, where $P(A|X,\lambda) = \mathrm{Tr}\big(\Pi_{(A|X)} \rho_{\mathcal{AB}}^{\lambda} \Pi_{(A|X)}^\dagger\big)$. In the process, Bob's reduced state updates from $\rho_{\mathcal{B}}^{\lambda} = \mathrm{Tr}_{\mathcal A}(\rho_{\mathcal AB}^{\lambda})$ to $\rho_{\mathcal B}^{(A|X,\lambda)} \equiv \mathrm{Tr}_{\mathcal A}(\rho_{\mathcal AB}^{(A|X,\lambda)})$. Bob's initial state $\rho_{\mathcal B}^{\lambda}$ encodes subjective probabilities $P(B|Y,\lambda) = \mathrm{Tr}(E_{B|Y} \rho_{\mathcal B}^{\lambda})$ for any measurement $Y$ on his local Hilbert space $\mathcal{H}_{\mathcal B}$. Upon learning the values ${A,X}$, Bob computes the conditional probabilities $P(B\mid A,X,Y,\lambda) = \mathrm{Tr}_{\mathcal B}\bigg[\Pi_{(B|Y)}\rho_{\mathcal B}^{(A|X,\lambda)}\bigg] = \frac{\mathrm{Tr}\bigl((\Pi_{(A|X)} \otimes \Pi_{(B|Y)}) \rho_{AB}^{\lambda}\bigr)}{P(A|X,\lambda)}$. This, in light of Eq. (\ref{eq:steerable}), has the form $ \frac{P(A,B|X,Y,\lambda)}{P(A|X,Y,\lambda)}$. Thus, the updated probability ${P(B \mid A,X,Y,\lambda)}$ encodes Bob's  probability assignment conditioned on new information ${A,X}$ via joint correlations in $\rho_{\mathcal AB}^{\lambda}$ in exactly the manner of a Bayesian update. On the strength of this, a subjective interpretation maintains that no physical evolution of Bob's local system occurs during the transition of $\rho_{\mathcal B}^{\lambda}$ to $\rho_{\mathcal B}^{(A|X,\lambda)}$---only informational conditioning on Alice's communicated outcome. Therefore, the dependence of Bob's conditional state on Alice's variables is considered as a Bayesian update. Evidently, such an interpretation is consistent with the assumption RC. On the other hand, by virtue of this same dependence, the distribution Eq. (\ref{eq:steerable}) lacks the factorizable form Eq. (\ref{eq:fp}). Thus, it is not consistent with FP. Referring to Table \ref{tab:classical_vs_quantum}, we find that such an interpretation entails a quantum causal structure and forbids a classical one.

By contrast, consider a realist interpretation of $P(A,B|X,Y)$ in Eq. (\ref{eq:steerable}). Here, by Definition \ref{def:realist}, the dependence of Bob's state on Alice's variables can't be considered as a Bayesian update.  Instead, the update of Bob's state is interpreted as an objective disturbance on the state of affairs at his side. This entails that in the absence of such remote disturbance, the dependence of Bob's state on Alice's measurement vanishes. In other words, in Eq. (\ref{eq:steerable}), the term $\rho_{A, X,\lambda}$ reduces to $\rho_{\lambda}$. Correspondingly, his state is left in the unsteerable assemblage:
\begin{equation}
    \tilde{\rho}^{A|X}_B = \sum_{\Lambda=\lambda}p(\lambda) P(A|X,\lambda) \tilde{\rho}_B^{\lambda}.
    \label{eq:nonsteer}
\end{equation}
Eq. (\ref{eq:nonsteer}) is consistent with the l.h.s of Eq. (\ref{eq:rp+rcs}) and thus incapable of producing a Bell's inequality violation.
Therefore, the dependence of Bob's state on Alice's variables in Eq. (\ref{eq:steerable}) would be interpreted in this interpretation as a violation of RC rather than of FP. (By contrast, it is the other way round in a subjective interpretation, as we noted above). Referring to Table \ref{tab:classical_vs_quantum}, we find that therefore a realist interpretation entails a classical causal structure and forbids a quantum one. As noted earlier, a classical causal model of Bell nonlocality would require Fine-Tuning to hide the superluminal signals at the macro-statistical level. 
\end{proof}

\begin{comment}
\begin{proposition} \color{red}
[\bf Subjective interpretation and quantum causal structure] In any given subjective interpretation of QM, a causal explanation for Bell nonlocality requires a quantum causal structure.
\label{thm:subj}
\end{proposition}

\begin{proof}
We know that the violation of Bell's inequality leaves Bob's state in the steered assemblage given by Eq. (\ref{eq:steer}). By Definition \ref{def:subjective}, the quantum state represents the observer's knowledge or belief about the system's state.
Therefore, the remote measurement disturbance produced by Alice constitutes an epistemic rather than \textcolor{red{What does it mean? real} causal influence. Thus, we aren't required to abandon the assumption of RC in Eq. (\ref{eq:rp+rcs}). However, FP must be relinquished since CC is assumed as a basic principle of the framework. This position (reject FP, uphold RC) corresponds to a quantum causal framework, as defined by Definition \ref{def:quantumc}), cf. Table \ref{tab:classical_vs_quantum}.
\end{proof}
\end{comment}

In retrospect, Proposition \ref{thm:onto} isn't surprising because a certain kind of classical worldview underlies both a realist interpretation as well as a classical causal model, whereas a willingness to abandon classical intuition underlies a subjective interpretation as well as a quantum causal model. Interestingly, certain nonclassical theories may inherently favor a realist interpretation and thus a classical causal model, e.g., Weinberg's nonlinear quantum theory \cite{weinberg1989testing}. However, this unambiguous interpretation of the quantum state comes at a price: that of allowing superluminal signaling (Appendix \ref{sec:weinberg}).    

\section{Quantum interpretation implications for causal structure of quantum computation \label{sec:1wqc}}
Upon some reflection, the question of reality of the quantum state naturally suggests the question of interpretation of quantum computation: Is a quantum computation an objective process happening in Nature, or is it an epistemic process happening in the observer's belief, expectation or information of the system's behavior? To be able to apply the causal considerations of the preceding section, we require to identify instances of Bell nonlocality in quantum computation. But as it happens, while the role of quantum entanglement plays in quantum speedup has been studied \cite{rungta2009quadratic}, that of quantum nonlocality has usually not. In the following two subsections, we will consider two instances where nonlocality plays a role in quantum computation, and consider their implications vis-\`a-vis the reality question. 

\subsection{Quantum computation, Bell nonlocality and causality \label{subsec:QCBell}}
It is known that Bell-type inequalities such as Eq. (\ref{eq:BI}) are derived from the operational assumptions of \textit{no-signaling} and \textit{predictability} \cite{cavalcanti2012bell}. Suppose instead we consider a modified set of assumptions, one where the former assumption remains, while we strengthen the latter assumption to (computationally) \textit{hard predictability}. By the latter term we mean that the prediction requires solving a problem for which no efficient classical algorithm is believed to exist, while there is a known efficient quantum algorithm.

Let us clarify this further. \textbf{BPP} is the set of problems that have efficient solutions on a randomized classical computer, whereas \textbf{BQP} is the class of problems having efficient solution on a quantum computer. In fact, it is not known whether factoring has an efficient classical randomized algorithm but it does indeed have an efficient quantum algorithm. Thus hard predictability refers to problems in \textbf{BQP} but believed to be outside \textbf{BPP}.

%\textcolor{green}{FOR EASY REFERENCE 2. Relation to Quantum Computation and Nonlocality: The second claimed contribution concerns quantum computation involving nonlocality. The authors introduce Assumptions 10 (no-signalling) and 11 (hard predictability), and then argue: “Suppose this Bell-type inequality is evaluated by local measurements after a polynomial-time evolution of the quantum computer, and found to be violated. Assumption 10 is guaranteed by ensuring that the measurements of different parties are performed near-simultaneously, necessitating abandoning Assumption 11.” However, this claim lacks convincing support. First, simultaneity of spatially separated events is frame-dependent. Second, it is unclear how signalling can be definitively ruled out in such scenarios. The authors further claim that “the classical causal influences required to violate the Bell inequality should not only be superluminal but also super-Turing.” Without stronger justification, the conclusion that super-Turing computation is required does not follow; superluminal communication alone could suffice.}

\begin{assumption}[\bf No-signalling] Conditional independence of output on the input of other parties:
\begin{equation}
    P(a | x, y, z,\cdots, \Gamma) = P(a | x, \Gamma),
    \label{eq:local}
\end{equation}
and analogously for other parties, where $\Gamma$ is the operational preparation information for the entangled state on which the measurements $x,y,z,\cdots$ are performed. 
\label{def:nosignaling}
\end{assumption}
\begin{assumption}
    [\bf Hard predictability] The conjunction of predictability and an assumption about complexity:
    \begin{enumerate}
        \item Predictability:
        \begin{equation}
       P(a,b,c, \cdots | x, y, z, \cdots, \Gamma) \in \{0, 1\},
       \label{eq:predict}
        \end{equation}
        meaning that specifying all inputs suffices to determine all outcomes deterministically.
        \item Hardness: $\Gamma$ represents information to efficiently prepare using a quantum computer a pure entangled $\ket{\Xi}$ on $n$ qubits, and there is no polynomial $r$ such that $\ket{\Xi}$ can be computed on a classical computer in $r(n)$ time. 
    \end{enumerate}
\label{def:hardpredict}
\end{assumption} 
By the hardness feature in Assumption \ref{def:hardpredict}, the entangled state specified by $\Gamma$ is assumed to be classically hard to compute. For example, it may require solving a system of linear equations, which can be performed efficiently by the HHL quantum algorithm \cite{harrow2009quantum}.

Consider testing a Bell-type inequality in the above setup. We will use the notation $y_{1,\cdots}$ to indicate $y_1,y_2,y_3,\cdots, y_n$ and $y_{m,\cdots}$ to indicate $y_m,y_{m+1},y_{m+2},\cdots, y_n$, where $m < n$. By recursive application of Bayesian inference, we have for any correlation 
\begin{subequations}
\begin{align}
   P&(a_{1,\cdots} | x_{1,\cdots}, \Gamma) = P(a_1 | a_{2,\cdots}, x_{1,\cdots}, \Gamma) \nonumber \\ &\times P(a_2 | a_{3,\cdots}, x_{1,\cdots},\Gamma) \cdots \times  P(a_n | x_{1,\cdots}, \Gamma). \label{eq:factora}\\
   &= P(a_1 | x_{1,\cdots}, \Gamma) P(a_2 | x_{1,\cdots},\Gamma) \times \cdots \times P(a_n | x_{1,\cdots}, \Gamma) \label{eq:factorb}\\
   &= P(a_1 | x_1, \Gamma) P(a_2 | x_2,\Gamma) \times \cdots \times P(a_n | x_n, \Gamma).
   \label{eq:factorc}
\end{align}
\label{eq:factor}
   \end{subequations}
Above, Eq. (\ref{eq:factorb}) follows from Eq. (\ref{eq:factora}) by virtue of assumption Eq. (\ref{eq:predict}), and Eq. (\ref{eq:factorc}) follows from Eq. (\ref{eq:factorb}) by the application of Eq. (\ref{eq:local}). The factorizable form of Eq. (\ref{eq:factor}) can be used to construct a Bell-type inequality $\mathfrak{B}$ \cite{cavalcanti2012bell}.

Suppose this Bell-type inequality is evaluated by local measurements after a polynomial-time evolution of the quantum computer, and found to be violated. Here we are given that Assumption \ref{def:nosignaling} is guaranteed by ensuring that the measurements of different parties are spacelike separated. Consider a realist interpretation applied to this quantum computational nonlocal scenario. Because the meausrements are spacelike separated, the classical causal influences required to violate the Bell inequality have to be superluminal. However if, in addition, the realist computational model is not super-Turing, then it will lack the resources to compute the values of $a_j$ to be communicated superluminally to produce the Bell's inequality violation. In other words, the realist model is required to \textit{superluminally} convey \textit{computationally hard} information. This augments the burden on the causal influences that must be hidden by Fine-Tuning, and thus seems to accentuate the Fine-Tuning problem.

By contrast, the subjective interpretation encounters a different kind of difficulty. Applying Proposition  \ref{thm:onto} to this quantum computational nonlocal scenario, a subjective interpretation conforms to the principle of RC and thus avoids superluminal causal influences. The entire burden of the nonclassical computation falls on the modification of the Reichenbach principle. This is somehow supposed to correspond to the evolution of the observer's belief, expectation or information of the system's behavior. If we conceptualize the human brain as a classical information processor with power in  $\mathcal{BPP}$, it seems unnatural that the dynamics of her subjective perception of such system can correspond to efficient solution of hard problems. (Of course, this difficulty holds even without considering the issue of Bell nonlocality.) The causal formulation of the quantum interpretation problem thus highlights the specific issues that inform the dilemma between realist and subjective interpretations of quantum mechanics.

\begin{comment}
\begin{figure}
    \centering
\begin{tikzpicture}[scale=1.5]

% Draw the circles (nodes)
\foreach \x/\y in {0/0, 1/0, 2/0, 3/0, 0/1, 1/1, 2/1, 3/1, 0/2, 1/2, 2/2, 3/2}
    \node[circle, draw, inner sep=2pt] (n\x\y) at (\x,\y) {};

% Draw the arrows
\draw[->, double] (n00) -- (n01);
\draw[->, double] (n01) -- (n02);
\draw[->, double] (n01) -- (n12);
\draw[->, double] (n00) -- (n11);
\draw[->, double] (n10) -- (n11);
\draw[->, double] (n11) -- (n12);
\draw[->, double] (n11) -- (n22);
\draw[->, double] (n11) -- (n21);
\draw[->, double] (n21) -- (n20);
\draw[->, double] (n21) -- (n30);
\draw[->, double] (n21) -- (n31);
\draw[->, double] (n21) -- (n32);

%\draw[->, thick] (n10) -- (n11);
%\draw[->, thick] (n11) -- (n12);
%\draw[->, thick] (n20) -- (n21);
%\draw[->, thick] (n21) -- (n22);
%\draw[->, thick] (n30) -- (n31);
%\draw[->, thick] (n31) -- (n32);

%\draw[->, thick] (n10) -- (n01);
%\draw[->, thick] (n20) -- (n11);
%\draw[->, thick] (n30) -- (n21);
%\draw[->, thick] (n22) -- (n11);
%\draw[->, thick] (n12) -- (n21);
%\draw[->, thick] (n02) -- (n11);
%\draw[->, thick] (n32) -- (n21);

%\draw[->, thick] (n11) -- (n21);
%\draw[->, thick] (n21) -- (n31);
%\draw[->, thick] (n11) -- (n01);
%\draw[->, thick] (n21) -- (n11);

\end{tikzpicture}
    \caption{Caption}
    \label{fig:QMA}
\end{figure}

\end{comment}

\subsection{Nonlocality of one-way quantum computation \label{subsec:1WQC}}
  
A one-way quantum computer (1WQC) provides a measurement-based model of universal quantum computation \cite{raussendorf2001one}. Instead of a sequence of unitaries (gates) as in standard quantum computation, this scheme utilizes single-qubit measurements applied on an entangled multi-qubit state called the \textit{cluster state}.  Given a lattice of qubits, a cluster state $\ket{\Psi_C}$ can be defined in terms of the eigenstate equations $\sigma_x^{(j)} \bigotimes_{k \in \mathscr{N}(j)} \sigma_z^{(k)}\ket{\Psi_C} = \pm\ket{\Psi_C}$, where $\mathscr{N}(j)$ denotes the neighborhood of the given qubit $j$ on the lattice. The single-qubit measurements are restricted to the basis $\mathscr{B}(\theta) = \{\frac{1}{\sqrt{2}}(\ket{0} \pm e^{i\theta}\ket{1})\}$.

In 1WQC, the computer program consists of the schedule of single-qubit measurements on the initial cluster state.  The information flow in a 1WQC happens by means of elementary quantum teleportation circuits \cite{bennett1993teleporting}, progressively destroying the initial entanglement. The random outcome of each single-qubit measurement determines the basis of the subsequent measurement. Each measured particle $D$ collapses the state of the remaining fraction of the cluster state, and the 1WQC performs computation by nonlocally teleporting quantum information to this fraction. The teleported quantum information employs two channels going from Alice to Bob: 
\begin{enumerate}
    \item A classical channel to transmit two bits.
    \item An Einstein-Podolsky-Rosen (EPR) channel to transfer the quantum state modulo a Pauli gate \cite{bennett1993teleporting}.
\end{enumerate}

\begin{comment}
As a basic illustration, to rotate a state $\psi$, one starts with the 5-qubit state $\ket{\psi}\ket{+,+,+,+}$ to which the cluster preparation entangling operation is applied, which yields $\frac{1}{2}\bigg[\ket{\psi}(\ket{0,-,0,-} + \ket{0,+,1,+}) + (\sigma_z\ket{\psi})(\ket{1,+,0,-} + \ket{1,-,1,+})\bigg]$. Denote the measurements on the first four qubits by $s_1, s_2, s_3, s_4 \in \{0,1\}$. Sequentially performing the measurements $\mathscr{B}(0), \mathscr{B}((-1)^{s_1 + 1}\theta_1), \mathscr{B}((-1)^{s_2}\theta_2), \mathscr{B}((-1)^{s_2+s_3}\theta_3)$ on qubits \#0, \#1, \#2, \#3 teleports the state $U_R(\theta_1, \theta_2, \theta_3)\ket{\psi}$ to qubit \#4, apart from a state-independent rotation $\sigma_z^{s_1+s_3}\sigma_x^{s_2+s_4}$ \cite{raussendorf2001one}. The  measurements cannot be performed simultaneously because later measurements depend on the outcomes of earlier ones. The C-NOT operation, similarly, can be implemented using only single qubit measurements, as indicated by the vertical shaded band in Fig. \ref{fig:QW}. 
\end{comment}

As a simple example, suppose we aim to implement the arbitrary single-qubit unitary:
\(
U(\alpha,\beta,\gamma) = R_z(\alpha) R_x(\beta) R_z(\gamma),
\)
where $\alpha,\beta,\gamma$ are the Euler angles, with \( R_z(\theta) = e^{-i\theta Z/2} \) (Z-rotation), and
 \( R_x(\theta) = e^{-i\theta X/2} \) (X-rotation).

The initial cluster state preparation preparation is by:
\(
\ket{C_3} = \text{CZ}_{2,3} \, \text{CZ}_{1,2} \, \big(\ket{\psi} \otimes \ket{+} \otimes \ket{+} \big).
\)
Define the measurement basis $M(\eta) = \{\frac{1}{\sqrt{2}}(\ket{0} \pm e^{i\eta}\ket{1})$.
The following adaptive measurement schedule is implemented:
(1) Qubit 1 is measured in the basis angle $M(\alpha)$ producing outcome $s_1 \in \{\pm1\}$. Qubit 2 is measured in the basis $M(s_1\beta)$ producing outcome $s_2$. Qubit 3 becomes the output barring a Pauli correction. The actual implemented unitary is:
        \(
        \ket{\text{out}} = X^{s_2} Z^{s_1} R_z(\alpha) R_x(\beta) R_z(\gamma) \ket{\psi}.
        \)

\begin{figure}
    \centering
\begin{quantikz}[row sep=0.4cm, column sep=0.5cm]
\lstick{$\ket{\psi}$} & \ctrl{1} & \qw        & \meter{$M(-\alpha)$} & \rstick[wires=1] {$s_1$} \\
\lstick{$\ket{+}$}    & \control{} & \ctrl{1} & \meter{$M(-\beta)$} & \rstick[wires=1]{$s_2$} \\
\lstick{$\ket{+}$}    & \qw      & \control{} & \qw & \gate[wires=1]{X^{s_2}Z^{s_1}R_z(\gamma)} \rstick{$U(\alpha,\beta,\gamma)\ket{\psi}$}
\end{quantikz}
    \caption{Quantum circuit with cluster state and feed-forward}
    \label{fig:enter-label}
\end{figure}

\begin{enumerate}
    \item Final state at qubit 3 becomes:
        \[
        \ket{\text{out}} = X^{s_2} Z^{s_1} R_z(\zeta) R_x(\eta) R_z(\xi) \ket{\psi}
        \]
        \item The byproduct operators \(X^{s_2}\), \(Z^{s_1}\) depend on measurement outcomes \(s_1, s_2 \in \{0,1\}\).
\end{enumerate}

Consider the application of a realist interpretation to the 1WQC. By proposition \ref{thm:onto}, this requires a classical causal model, whose corresponding causal structure is depicted in Fig. \ref{fig:placeholder}.  It is worth noting that this causal structure involves two types of causal influences, corresponding to the dual classical and EPR channels of quantum teleportation: 
\begin{description}
    \item[Single arrows] representing temporally sequential, adaptive measurements, mediated by the transmission of classical information.
    \item[Double arrows] representing the superluminal causal influence corresponding to an EPR channel \cite{bennett1993teleporting}.
\end{description} 
The two types of arrows play a complementary role in the DAG: the latter teleports the quantum state modulo a discrete rotation, while the former is necessary to enforce no-signaling. 

Suppose the 1WQC is being used to run Shor's or the HLL algorithm. Here again the fact that the transferred information is sufficient to solve a classically intractable problem augments the burden on the causal influences that must be hidden by Fine-Tuning, and thus (as in Section \ref{subsec:QCBell}) seems to accentuate the Fine-Tuning problem.

\begin{figure}
    \centering
    \begin{tikzpicture}[scale=1.4]
	% Define quantum states and transition points
	\node (psi) at (0,0) {$|\psi\rangle$};
	\node[circle, draw, inner sep=2pt] (c1) at (2,0) {};
	\node[circle, draw, inner sep=2pt] (c2) at (4,0) {};
	\node (Upsi) at (4.8,0) {$U(\alpha, \beta, \gamma)|\psi\rangle$};
	
	% Curved double arrows from psi to c1 and c1 to c2
	\draw[->, double, blue] (psi) to[out=45, in=135] (c1);
	\draw[->, double, blue] (c1) to[out=45, in=135] (c2);
	
	% Single black arrows
	\draw[->] (psi) to[out=-45, in=225] node[midway, below] {$s_1$} (c1);
	\draw[->] (c1) to[out=-45, in=225] node[midway, below] {$s_2$} (c2);
\end{tikzpicture}
   \caption{Realist interpretation of 1WQC employing a classical causal model of the 1WQC in Fig. \ref{fig:enter-label}: it involves two types of causal influences, mirroring the dual channels of quantum teleportation: (a) Conventional causal influences, indicated by single arrows: they correspond to classical information communicated to update the temporal sequence of adaptive measurements; (b) Superluminal causal influences, indicated by double-arrows: they correspond to the EPR channel of teleportation.}
    \label{fig:placeholder}
\end{figure}

\section{\label{sec:discus} Discussion and conclusions } 

Various frameworks have been developed in order to elucidate the mathematical structure of QM and identify the broad physical principles in Nature that single out quantum mechanics as special in ``theoryspace''. Among these are the framework of generalized probability theories (GPTs) \cite{barrett2007information, MR3241077, sorkin1994quantum, lee2015computation}, general physical theories \cite{lee2016deriving, lee2016bounds}, derivation of QM from information theoretic principles \cite{d2017quantum, chiribella2010probabilistic, chiribella2011informational}, and other approaches which study ad hoc modifications of QM \cite{abrams1998nonlinear, aaronson2004quantum, aaronson2005quantum}. Can these different frameworks (GPT, etc.) that shed light on the quantum formalism, throw any light on the quantum measurement or interpretation problem?  Unfortunately not, and what is more, these problems afflict theories in those frameworks also \cite{baumann2021generalized}. However, the problems are usually ignored, possibly because these theories are operational and hypothetical. An exception here would be Weinberg's nonlinear QM, wherein the quantum state naturally favors the realist interpretation in that theory. However, this definitiveness comes at the price of superluminal signaling. This brings up the question of whether a natural or reasonable nonclassical theory could provide a resolution to its own interpretation question. An argument for an affirmative answer to this question was presented in Ref. \cite{srikanth2022operational}.

The present work initiates the prospect of applying causality and computational complexity to quantum interpretational problems. One possible future direction would be to develop a complexity theoretically and causally informed interpretation of QM. Going farther, one may construct a computational complexity and/or causal structure inspired ontological framework for GPTs, which uses resource constraints interwoven into the axioms of operational theories. For example, the concept of affine Turing machines have been used to characterize the computational landscape of the GPTs \cite{barrett2019computational}. In turn, this suggests the possibility of a ``landscape of physical theories" derived from computational models \cite{kumar2025on}. Also, the implications of our present work for the Many-Worlds Interpretation (MWI) \cite{dewitt2015many} may be explored. By Proposition \ref{thm:onto}, as a realist interpretation, MWI is characterized by a natural preference for a classical causal structure. However, this Proposition implicitly assumes a single universe, and its extension to accommodate the concepts of a classical or nonclassical causal structure in a multiverse scenario needs to established.

\section*{Funding information}
RS was partially supported by the Indian Science \& Engineering Research Board (SERB) grant CRG/2022/008345.

%\bibliography{apssamp}% Produces the bibliography via BibTeX.

\begin{thebibliography}{60}%
\makeatletter
\providecommand \@ifxundefined [1]{%
 \@ifx{#1\undefined}
}%
\providecommand \@ifnum [1]{%
 \ifnum #1\expandafter \@firstoftwo
 \else \expandafter \@secondoftwo
 \fi
}%
\providecommand \@ifx [1]{%
 \ifx #1\expandafter \@firstoftwo
 \else \expandafter \@secondoftwo
 \fi
}%
\providecommand \natexlab [1]{#1}%
\providecommand \enquote  [1]{``#1''}%
\providecommand \bibnamefont  [1]{#1}%
\providecommand \bibfnamefont [1]{#1}%
\providecommand \citenamefont [1]{#1}%
\providecommand \href@noop [0]{\@secondoftwo}%
\providecommand \href [0]{\begingroup \@sanitize@url \@href}%
\providecommand \@href[1]{\@@startlink{#1}\@@href}%
\providecommand \@@href[1]{\endgroup#1\@@endlink}%
\providecommand \@sanitize@url [0]{\catcode `\\12\catcode `\$12\catcode `\&12\catcode `\#12\catcode `\^12\catcode `\_12\catcode `\%12\relax}%
\providecommand \@@startlink[1]{}%
\providecommand \@@endlink[0]{}%
\providecommand \url  [0]{\begingroup\@sanitize@url \@url }%
\providecommand \@url [1]{\endgroup\@href {#1}{\urlprefix }}%
\providecommand \urlprefix  [0]{URL }%
\providecommand \Eprint [0]{\href }%
\providecommand \doibase [0]{http://dx.doi.org/}%
\providecommand \selectlanguage [0]{\@gobble}%
\providecommand \bibinfo  [0]{\@secondoftwo}%
\providecommand \bibfield  [0]{\@secondoftwo}%
\providecommand \translation [1]{[#1]}%
\providecommand \BibitemOpen [0]{}%
\providecommand \bibitemStop [0]{}%
\providecommand \bibitemNoStop [0]{.\EOS\space}%
\providecommand \EOS [0]{\spacefactor3000\relax}%
\providecommand \BibitemShut  [1]{\csname bibitem#1\endcsname}%
\let\auto@bib@innerbib\@empty
%</preamble>
\bibitem [{\citenamefont {Cabello}(2017)}]{Cabello_2017}%
  \BibitemOpen
  \bibfield  {author} {\bibinfo {author} {\bibfnamefont {A.}~\bibnamefont {Cabello}},\ }\enquote {\bibinfo {title} {Interpretations of quantum theory: A map of madness},}\ in\ \href@noop {} {\emph {\bibinfo {booktitle} {What is Quantum Information?}}},\ \bibinfo {editor} {edited by\ \bibinfo {editor} {\bibfnamefont {O.}~\bibnamefont {Lombardi}}, \bibinfo {editor} {\bibfnamefont {S.}~\bibnamefont {Fortin}}, \bibinfo {editor} {\bibfnamefont {F.}~\bibnamefont {Holik}}, \ and\ \bibinfo {editor} {\bibfnamefont {C.}~\bibnamefont {López}}}\ (\bibinfo  {publisher} {Cambridge University Press},\ \bibinfo {year} {2017})\ p.\ \bibinfo {pages} {138–144}\BibitemShut {NoStop}%
\bibitem [{\citenamefont {Leifer}(2014)}]{leifer2014quantum}%
  \BibitemOpen
  \bibfield  {author} {\bibinfo {author} {\bibfnamefont {M.~S.}\ \bibnamefont {Leifer}},\ }\href@noop {} {\bibfield  {journal} {\bibinfo  {journal} {arXiv preprint arXiv:1409.1570}\ } (\bibinfo {year} {2014})}\BibitemShut {NoStop}%
\bibitem [{\citenamefont {Bassi}\ \emph {et~al.}(2021)\citenamefont {Bassi} \emph {et~al.}}]{bassi2021philosophy}%
  \BibitemOpen
  \bibfield  {author} {\bibinfo {author} {\bibfnamefont {A.}~\bibnamefont {Bassi}} \emph {et~al.},\ }\href@noop {} {\bibfield  {journal} {\bibinfo  {journal} {Oxford Research Encyclopedias}\ } (\bibinfo {year} {2021})}\BibitemShut {NoStop}%
\bibitem [{\citenamefont {Ghirardi}\ \emph {et~al.}(1986)\citenamefont {Ghirardi}, \citenamefont {Rimini},\ and\ \citenamefont {Weber}}]{ghirardi1986unified}%
  \BibitemOpen
  \bibfield  {author} {\bibinfo {author} {\bibfnamefont {G.~C.}\ \bibnamefont {Ghirardi}}, \bibinfo {author} {\bibfnamefont {A.}~\bibnamefont {Rimini}}, \ and\ \bibinfo {author} {\bibfnamefont {T.}~\bibnamefont {Weber}},\ }\href@noop {} {\bibfield  {journal} {\bibinfo  {journal} {Physical review D}\ }\textbf {\bibinfo {volume} {34}},\ \bibinfo {pages} {470} (\bibinfo {year} {1986})}\BibitemShut {NoStop}%
\bibitem [{\citenamefont {Di{\'o}si}(1989)}]{diosi1989models}%
  \BibitemOpen
  \bibfield  {author} {\bibinfo {author} {\bibfnamefont {L.}~\bibnamefont {Di{\'o}si}},\ }\href@noop {} {\bibfield  {journal} {\bibinfo  {journal} {Physical Review A}\ }\textbf {\bibinfo {volume} {40}},\ \bibinfo {pages} {1165} (\bibinfo {year} {1989})}\BibitemShut {NoStop}%
\bibitem [{\citenamefont {Penrose}(1996)}]{penrose1996gravity}%
  \BibitemOpen
  \bibfield  {author} {\bibinfo {author} {\bibfnamefont {R.}~\bibnamefont {Penrose}},\ }\href@noop {} {\bibfield  {journal} {\bibinfo  {journal} {General relativity and gravitation}\ }\textbf {\bibinfo {volume} {28}},\ \bibinfo {pages} {581} (\bibinfo {year} {1996})}\BibitemShut {NoStop}%
\bibitem [{\citenamefont {Patel}\ and\ \citenamefont {Kumar}(2017)}]{patel2017weak}%
  \BibitemOpen
  \bibfield  {author} {\bibinfo {author} {\bibfnamefont {A.}~\bibnamefont {Patel}}\ and\ \bibinfo {author} {\bibfnamefont {P.}~\bibnamefont {Kumar}},\ }\href@noop {} {\bibfield  {journal} {\bibinfo  {journal} {Physical Review A}\ }\textbf {\bibinfo {volume} {96}},\ \bibinfo {pages} {022108} (\bibinfo {year} {2017})}\BibitemShut {NoStop}%
\bibitem [{\citenamefont {Goldstein}(2001)}]{goldstein2001bohmian}%
  \BibitemOpen
  \bibfield  {author} {\bibinfo {author} {\bibfnamefont {S.}~\bibnamefont {Goldstein}},\ }\href@noop {} {\  (\bibinfo {year} {2001})}\BibitemShut {NoStop}%
\bibitem [{\citenamefont {Vaidman}(2002)}]{vaidman2002many}%
  \BibitemOpen
  \bibfield  {author} {\bibinfo {author} {\bibfnamefont {L.}~\bibnamefont {Vaidman}},\ }\href@noop {} {\  (\bibinfo {year} {2002})}\BibitemShut {NoStop}%
\bibitem [{\citenamefont {Dewitt}\ and\ \citenamefont {Graham}(2015)}]{dewitt2015many}%
  \BibitemOpen
  \bibfield  {author} {\bibinfo {author} {\bibfnamefont {B.~S.}\ \bibnamefont {Dewitt}}\ and\ \bibinfo {author} {\bibfnamefont {N.}~\bibnamefont {Graham}},\ }\href@noop {} {\emph {\bibinfo {title} {The many-worlds interpretation of quantum mechanics}}},\ Vol.~\bibinfo {volume} {63}\ (\bibinfo  {publisher} {Princeton University Press},\ \bibinfo {year} {2015})\BibitemShut {NoStop}%
\bibitem [{\citenamefont {Spekkens}(2007)}]{spekkens2007evidence}%
  \BibitemOpen
  \bibfield  {author} {\bibinfo {author} {\bibfnamefont {R.~W.}\ \bibnamefont {Spekkens}},\ }\href@noop {} {\bibfield  {journal} {\bibinfo  {journal} {Physical Review A}\ }\textbf {\bibinfo {volume} {75}},\ \bibinfo {pages} {032110} (\bibinfo {year} {2007})}\BibitemShut {NoStop}%
\bibitem [{\citenamefont {Faye}(2002)}]{faye2002copenhagen}%
  \BibitemOpen
  \bibfield  {author} {\bibinfo {author} {\bibfnamefont {J.}~\bibnamefont {Faye}},\ }\href@noop {} {\  (\bibinfo {year} {2002})}\BibitemShut {NoStop}%
\bibitem [{\citenamefont {Fuchs}\ \emph {et~al.}(2014)\citenamefont {Fuchs}, \citenamefont {Mermin},\ and\ \citenamefont {Schack}}]{fuchs2014introduction}%
  \BibitemOpen
  \bibfield  {author} {\bibinfo {author} {\bibfnamefont {C.~A.}\ \bibnamefont {Fuchs}}, \bibinfo {author} {\bibfnamefont {N.~D.}\ \bibnamefont {Mermin}}, \ and\ \bibinfo {author} {\bibfnamefont {R.}~\bibnamefont {Schack}},\ }\href@noop {} {\bibfield  {journal} {\bibinfo  {journal} {American Journal of Physics}\ }\textbf {\bibinfo {volume} {82}},\ \bibinfo {pages} {749} (\bibinfo {year} {2014})}\BibitemShut {NoStop}%
\bibitem [{\citenamefont {Fuchs}\ and\ \citenamefont {Peres}(2000)}]{fuchs2000quantum}%
  \BibitemOpen
  \bibfield  {author} {\bibinfo {author} {\bibfnamefont {C.~A.}\ \bibnamefont {Fuchs}}\ and\ \bibinfo {author} {\bibfnamefont {A.}~\bibnamefont {Peres}},\ }\href@noop {} {\bibfield  {journal} {\bibinfo  {journal} {Physics today}\ }\textbf {\bibinfo {volume} {53}},\ \bibinfo {pages} {70} (\bibinfo {year} {2000})}\BibitemShut {NoStop}%
\bibitem [{\citenamefont {Zeilinger}(1999)}]{zeilinger1999foundational}%
  \BibitemOpen
  \bibfield  {author} {\bibinfo {author} {\bibfnamefont {A.}~\bibnamefont {Zeilinger}},\ }\href@noop {} {\bibfield  {journal} {\bibinfo  {journal} {Foundations of Physics}\ }\textbf {\bibinfo {volume} {29}},\ \bibinfo {pages} {631} (\bibinfo {year} {1999})}\BibitemShut {NoStop}%
\bibitem [{\citenamefont {Brukner}(2017)}]{brukner2017quantum}%
  \BibitemOpen
  \bibfield  {author} {\bibinfo {author} {\bibfnamefont {{\v{C}}.}~\bibnamefont {Brukner}},\ }\href@noop {} {\bibfield  {journal} {\bibinfo  {journal} {Quantum [un] speakables II: half a century of Bell's theorem}\ ,\ \bibinfo {pages} {95}} (\bibinfo {year} {2017})}\BibitemShut {NoStop}%
\bibitem [{\citenamefont {Pearl}(2009)}]{pearl2009causality}%
  \BibitemOpen
  \bibfield  {author} {\bibinfo {author} {\bibfnamefont {J.}~\bibnamefont {Pearl}},\ }\href@noop {} {\emph {\bibinfo {title} {Causality}}}\ (\bibinfo  {publisher} {Cambridge university press},\ \bibinfo {year} {2009})\BibitemShut {NoStop}%
\bibitem [{\citenamefont {Reichenbach}(1991)}]{reichenbach1991direction}%
  \BibitemOpen
  \bibfield  {author} {\bibinfo {author} {\bibfnamefont {H.}~\bibnamefont {Reichenbach}},\ }\href@noop {} {\emph {\bibinfo {title} {The direction of time}}},\ Vol.~\bibinfo {volume} {65}\ (\bibinfo  {publisher} {Univ of California Press},\ \bibinfo {year} {1991})\BibitemShut {NoStop}%
\bibitem [{\citenamefont {Wood}\ and\ \citenamefont {Spekkens}(2015)}]{wood2015lesson}%
  \BibitemOpen
  \bibfield  {author} {\bibinfo {author} {\bibfnamefont {C.~J.}\ \bibnamefont {Wood}}\ and\ \bibinfo {author} {\bibfnamefont {R.~W.}\ \bibnamefont {Spekkens}},\ }\href@noop {} {\bibfield  {journal} {\bibinfo  {journal} {New Journal of Physics}\ }\textbf {\bibinfo {volume} {17}},\ \bibinfo {pages} {033002} (\bibinfo {year} {2015})}\BibitemShut {NoStop}%
\bibitem [{\citenamefont {Cavalcanti}(2018)}]{cavalcanti2018classical}%
  \BibitemOpen
  \bibfield  {author} {\bibinfo {author} {\bibfnamefont {E.~G.}\ \bibnamefont {Cavalcanti}},\ }\href@noop {} {\bibfield  {journal} {\bibinfo  {journal} {Physical Review X}\ }\textbf {\bibinfo {volume} {8}},\ \bibinfo {pages} {021018} (\bibinfo {year} {2018})}\BibitemShut {NoStop}%
\bibitem [{\citenamefont {Cavalcanti}\ and\ \citenamefont {Lal}(2014)}]{cavalcanti2014modifications}%
  \BibitemOpen
  \bibfield  {author} {\bibinfo {author} {\bibfnamefont {E.~G.}\ \bibnamefont {Cavalcanti}}\ and\ \bibinfo {author} {\bibfnamefont {R.}~\bibnamefont {Lal}},\ }\href@noop {} {\bibfield  {journal} {\bibinfo  {journal} {Journal of Physics A: Mathematical and Theoretical}\ }\textbf {\bibinfo {volume} {47}},\ \bibinfo {pages} {424018} (\bibinfo {year} {2014})}\BibitemShut {NoStop}%
\bibitem [{\citenamefont {Cavalcanti}\ and\ \citenamefont {Wiseman}(2021)}]{cavalcanti2021implications}%
  \BibitemOpen
  \bibfield  {author} {\bibinfo {author} {\bibfnamefont {E.~G.}\ \bibnamefont {Cavalcanti}}\ and\ \bibinfo {author} {\bibfnamefont {H.~M.}\ \bibnamefont {Wiseman}},\ }\href@noop {} {\bibfield  {journal} {\bibinfo  {journal} {Entropy}\ }\textbf {\bibinfo {volume} {23}},\ \bibinfo {pages} {925} (\bibinfo {year} {2021})}\BibitemShut {NoStop}%
\bibitem [{\citenamefont {Allen}\ \emph {et~al.}(2017)\citenamefont {Allen}, \citenamefont {Barrett}, \citenamefont {Horsman}, \citenamefont {Lee},\ and\ \citenamefont {Spekkens}}]{allen2017quantum}%
  \BibitemOpen
  \bibfield  {author} {\bibinfo {author} {\bibfnamefont {J.-M.~A.}\ \bibnamefont {Allen}}, \bibinfo {author} {\bibfnamefont {J.}~\bibnamefont {Barrett}}, \bibinfo {author} {\bibfnamefont {D.~C.}\ \bibnamefont {Horsman}}, \bibinfo {author} {\bibfnamefont {C.~M.}\ \bibnamefont {Lee}}, \ and\ \bibinfo {author} {\bibfnamefont {R.~W.}\ \bibnamefont {Spekkens}},\ }\href@noop {} {\bibfield  {journal} {\bibinfo  {journal} {Physical Review X}\ }\textbf {\bibinfo {volume} {7}},\ \bibinfo {pages} {031021} (\bibinfo {year} {2017})}\BibitemShut {NoStop}%
\bibitem [{\citenamefont {Coiteux-Roy}\ \emph {et~al.}(2021)\citenamefont {Coiteux-Roy}, \citenamefont {Wolfe},\ and\ \citenamefont {Renou}}]{coiteux2021no}%
  \BibitemOpen
  \bibfield  {author} {\bibinfo {author} {\bibfnamefont {X.}~\bibnamefont {Coiteux-Roy}}, \bibinfo {author} {\bibfnamefont {E.}~\bibnamefont {Wolfe}}, \ and\ \bibinfo {author} {\bibfnamefont {M.-O.}\ \bibnamefont {Renou}},\ }\href@noop {} {\bibfield  {journal} {\bibinfo  {journal} {Physical review letters}\ }\textbf {\bibinfo {volume} {127}},\ \bibinfo {pages} {200401} (\bibinfo {year} {2021})}\BibitemShut {NoStop}%
\bibitem [{\citenamefont {Lauand}\ \emph {et~al.}(2023)\citenamefont {Lauand}, \citenamefont {Poderini}, \citenamefont {Nery}, \citenamefont {Moreno}, \citenamefont {Pollyceno}, \citenamefont {Rabelo},\ and\ \citenamefont {Chaves}}]{lauand2023witnessing}%
  \BibitemOpen
  \bibfield  {author} {\bibinfo {author} {\bibfnamefont {P.}~\bibnamefont {Lauand}}, \bibinfo {author} {\bibfnamefont {D.}~\bibnamefont {Poderini}}, \bibinfo {author} {\bibfnamefont {R.}~\bibnamefont {Nery}}, \bibinfo {author} {\bibfnamefont {G.}~\bibnamefont {Moreno}}, \bibinfo {author} {\bibfnamefont {L.}~\bibnamefont {Pollyceno}}, \bibinfo {author} {\bibfnamefont {R.}~\bibnamefont {Rabelo}}, \ and\ \bibinfo {author} {\bibfnamefont {R.}~\bibnamefont {Chaves}},\ }\href@noop {} {\bibfield  {journal} {\bibinfo  {journal} {PRX Quantum}\ }\textbf {\bibinfo {volume} {4}},\ \bibinfo {pages} {020311} (\bibinfo {year} {2023})}\BibitemShut {NoStop}%
\bibitem [{\citenamefont {Navascu{\'e}s}\ and\ \citenamefont {Wolfe}(2020)}]{navascues2020inflation}%
  \BibitemOpen
  \bibfield  {author} {\bibinfo {author} {\bibfnamefont {M.}~\bibnamefont {Navascu{\'e}s}}\ and\ \bibinfo {author} {\bibfnamefont {E.}~\bibnamefont {Wolfe}},\ }\href@noop {} {\bibfield  {journal} {\bibinfo  {journal} {Journal of Causal Inference}\ }\textbf {\bibinfo {volume} {8}},\ \bibinfo {pages} {70} (\bibinfo {year} {2020})}\BibitemShut {NoStop}%
\bibitem [{\citenamefont {Barrett}\ \emph {et~al.}(2021)\citenamefont {Barrett}, \citenamefont {Lorenz},\ and\ \citenamefont {Oreshkov}}]{barrett2021cyclic}%
  \BibitemOpen
  \bibfield  {author} {\bibinfo {author} {\bibfnamefont {J.}~\bibnamefont {Barrett}}, \bibinfo {author} {\bibfnamefont {R.}~\bibnamefont {Lorenz}}, \ and\ \bibinfo {author} {\bibfnamefont {O.}~\bibnamefont {Oreshkov}},\ }\href@noop {} {\bibfield  {journal} {\bibinfo  {journal} {Nature communications}\ }\textbf {\bibinfo {volume} {12}},\ \bibinfo {pages} {885} (\bibinfo {year} {2021})}\BibitemShut {NoStop}%
\bibitem [{\citenamefont {Bong}\ \emph {et~al.}(2020)\citenamefont {Bong}, \citenamefont {Utreras-Alarc{\'o}n}, \citenamefont {Ghafari}, \citenamefont {Liang}, \citenamefont {Tischler}, \citenamefont {Cavalcanti}, \citenamefont {Pryde},\ and\ \citenamefont {Wiseman}}]{bong2020strong}%
  \BibitemOpen
  \bibfield  {author} {\bibinfo {author} {\bibfnamefont {K.-W.}\ \bibnamefont {Bong}}, \bibinfo {author} {\bibfnamefont {A.}~\bibnamefont {Utreras-Alarc{\'o}n}}, \bibinfo {author} {\bibfnamefont {F.}~\bibnamefont {Ghafari}}, \bibinfo {author} {\bibfnamefont {Y.-C.}\ \bibnamefont {Liang}}, \bibinfo {author} {\bibfnamefont {N.}~\bibnamefont {Tischler}}, \bibinfo {author} {\bibfnamefont {E.~G.}\ \bibnamefont {Cavalcanti}}, \bibinfo {author} {\bibfnamefont {G.~J.}\ \bibnamefont {Pryde}}, \ and\ \bibinfo {author} {\bibfnamefont {H.~M.}\ \bibnamefont {Wiseman}},\ }\href@noop {} {\bibfield  {journal} {\bibinfo  {journal} {Nature Physics}\ }\textbf {\bibinfo {volume} {16}},\ \bibinfo {pages} {1199} (\bibinfo {year} {2020})}\BibitemShut {NoStop}%
\bibitem [{\citenamefont {Wiseman}\ \emph {et~al.}(2023)\citenamefont {Wiseman}, \citenamefont {Cavalcanti},\ and\ \citenamefont {Rieffel}}]{wiseman2023thoughtful}%
  \BibitemOpen
  \bibfield  {author} {\bibinfo {author} {\bibfnamefont {H.~M.}\ \bibnamefont {Wiseman}}, \bibinfo {author} {\bibfnamefont {E.~G.}\ \bibnamefont {Cavalcanti}}, \ and\ \bibinfo {author} {\bibfnamefont {E.~G.}\ \bibnamefont {Rieffel}},\ }\href@noop {} {\bibfield  {journal} {\bibinfo  {journal} {Quantum}\ }\textbf {\bibinfo {volume} {7}},\ \bibinfo {pages} {1112} (\bibinfo {year} {2023})}\BibitemShut {NoStop}%
\bibitem [{\citenamefont {Ormrod}\ and\ \citenamefont {Barrett}(2024)}]{ormrod2024quantum}%
  \BibitemOpen
  \bibfield  {author} {\bibinfo {author} {\bibfnamefont {N.}~\bibnamefont {Ormrod}}\ and\ \bibinfo {author} {\bibfnamefont {J.}~\bibnamefont {Barrett}},\ }\href@noop {} {\bibfield  {journal} {\bibinfo  {journal} {arXiv preprint arXiv:2401.18005}\ } (\bibinfo {year} {2024})}\BibitemShut {NoStop}%
\bibitem [{\citenamefont {Liu}\ \emph {et~al.}(2025)\citenamefont {Liu}, \citenamefont {Qiu}, \citenamefont {Dahlsten},\ and\ \citenamefont {Vedral}}]{liu2025quantum}%
  \BibitemOpen
  \bibfield  {author} {\bibinfo {author} {\bibfnamefont {X.}~\bibnamefont {Liu}}, \bibinfo {author} {\bibfnamefont {Y.}~\bibnamefont {Qiu}}, \bibinfo {author} {\bibfnamefont {O.}~\bibnamefont {Dahlsten}}, \ and\ \bibinfo {author} {\bibfnamefont {V.}~\bibnamefont {Vedral}},\ }\href@noop {} {\bibfield  {journal} {\bibinfo  {journal} {npj Quantum Information}\ }\textbf {\bibinfo {volume} {11}},\ \bibinfo {pages} {1} (\bibinfo {year} {2025})}\BibitemShut {NoStop}%
\bibitem [{\citenamefont {Bell}(1976)}]{bell1976theory}%
  \BibitemOpen
  \bibfield  {author} {\bibinfo {author} {\bibfnamefont {J.}~\bibnamefont {Bell}},\ }\href@noop {} {\bibfield  {journal} {\bibinfo  {journal} {Epistemological Letters}\ }\textbf {\bibinfo {volume} {9}},\ \bibinfo {pages} {11} (\bibinfo {year} {1976})}\BibitemShut {NoStop}%
\bibitem [{\citenamefont {Bell}(1964)}]{bell1964on}%
  \BibitemOpen
  \bibfield  {author} {\bibinfo {author} {\bibfnamefont {J.}~\bibnamefont {Bell}},\ }\href@noop {} {\bibfield  {journal} {\bibinfo  {journal} {Physics}\ }\textbf {\bibinfo {volume} {1}},\ \bibinfo {pages} {195} (\bibinfo {year} {1964})}\BibitemShut {NoStop}%
\bibitem [{\citenamefont {Clauser}\ \emph {et~al.}(1969)\citenamefont {Clauser}, \citenamefont {Horne}, \citenamefont {Shimony},\ and\ \citenamefont {Holt}}]{clauser1969proposed}%
  \BibitemOpen
  \bibfield  {author} {\bibinfo {author} {\bibfnamefont {J.~F.}\ \bibnamefont {Clauser}}, \bibinfo {author} {\bibfnamefont {M.~A.}\ \bibnamefont {Horne}}, \bibinfo {author} {\bibfnamefont {A.}~\bibnamefont {Shimony}}, \ and\ \bibinfo {author} {\bibfnamefont {R.~A.}\ \bibnamefont {Holt}},\ }\href {\doibase 10.1103/PhysRevLett.23.880} {\bibfield  {journal} {\bibinfo  {journal} {Phys. Rev. Lett.}\ }\textbf {\bibinfo {volume} {23}},\ \bibinfo {pages} {880} (\bibinfo {year} {1969})}\BibitemShut {NoStop}%
\bibitem [{\citenamefont {Barrett}(2007)}]{barrett2007information}%
  \BibitemOpen
  \bibfield  {author} {\bibinfo {author} {\bibfnamefont {J.}~\bibnamefont {Barrett}},\ }\href@noop {} {\bibfield  {journal} {\bibinfo  {journal} {Physical Review A}\ }\textbf {\bibinfo {volume} {75}},\ \bibinfo {pages} {032304} (\bibinfo {year} {2007})}\BibitemShut {NoStop}%
\bibitem [{\citenamefont {Srikanth}()}]{srikanth2025}%
  \BibitemOpen
  \bibfield  {author} {\bibinfo {author} {\bibfnamefont {R.}~\bibnamefont {Srikanth}},\ }\href@noop {} {}\bibinfo {note} {Work under preparation}\BibitemShut {NoStop}%
\bibitem [{\citenamefont {Harrigan}\ and\ \citenamefont {Spekkens}(2010)}]{harrigan2010einstein}%
  \BibitemOpen
  \bibfield  {author} {\bibinfo {author} {\bibfnamefont {N.}~\bibnamefont {Harrigan}}\ and\ \bibinfo {author} {\bibfnamefont {R.~W.}\ \bibnamefont {Spekkens}},\ }\href {\doibase 10.1007/s10701-009-9347-0} {\bibfield  {journal} {\bibinfo  {journal} {Foundations of Physics}\ }\textbf {\bibinfo {volume} {40}},\ \bibinfo {pages} {125} (\bibinfo {year} {2010})}\BibitemShut {NoStop}%
\bibitem [{\citenamefont {Weinberg}(1989)}]{weinberg1989testing}%
  \BibitemOpen
  \bibfield  {author} {\bibinfo {author} {\bibfnamefont {S.}~\bibnamefont {Weinberg}},\ }\href@noop {} {\bibfield  {journal} {\bibinfo  {journal} {Annals of Physics}\ }\textbf {\bibinfo {volume} {194}},\ \bibinfo {pages} {336} (\bibinfo {year} {1989})}\BibitemShut {NoStop}%
\bibitem [{\citenamefont {Rungta}(2009)}]{rungta2009quadratic}%
  \BibitemOpen
  \bibfield  {author} {\bibinfo {author} {\bibfnamefont {P.}~\bibnamefont {Rungta}},\ }\href@noop {} {\bibfield  {journal} {\bibinfo  {journal} {Physics Letters A}\ }\textbf {\bibinfo {volume} {373}},\ \bibinfo {pages} {2652} (\bibinfo {year} {2009})}\BibitemShut {NoStop}%
\bibitem [{\citenamefont {Cavalcanti}\ and\ \citenamefont {Wiseman}(2012)}]{cavalcanti2012bell}%
  \BibitemOpen
  \bibfield  {author} {\bibinfo {author} {\bibfnamefont {E.~G.}\ \bibnamefont {Cavalcanti}}\ and\ \bibinfo {author} {\bibfnamefont {H.~M.}\ \bibnamefont {Wiseman}},\ }\href@noop {} {\bibfield  {journal} {\bibinfo  {journal} {Foundations of Physics}\ }\textbf {\bibinfo {volume} {42}},\ \bibinfo {pages} {1329} (\bibinfo {year} {2012})}\BibitemShut {NoStop}%
\bibitem [{\citenamefont {Harrow}\ \emph {et~al.}(2009)\citenamefont {Harrow}, \citenamefont {Hassidim},\ and\ \citenamefont {Lloyd}}]{harrow2009quantum}%
  \BibitemOpen
  \bibfield  {author} {\bibinfo {author} {\bibfnamefont {A.~W.}\ \bibnamefont {Harrow}}, \bibinfo {author} {\bibfnamefont {A.}~\bibnamefont {Hassidim}}, \ and\ \bibinfo {author} {\bibfnamefont {S.}~\bibnamefont {Lloyd}},\ }\href {\doibase 10.1103/PhysRevLett.103.150502} {\bibfield  {journal} {\bibinfo  {journal} {Phys. Rev. Lett.}\ }\textbf {\bibinfo {volume} {103}},\ \bibinfo {pages} {150502} (\bibinfo {year} {2009})}\BibitemShut {NoStop}%
\bibitem [{\citenamefont {Raussendorf}\ and\ \citenamefont {Briegel}(2001)}]{raussendorf2001one}%
  \BibitemOpen
  \bibfield  {author} {\bibinfo {author} {\bibfnamefont {R.}~\bibnamefont {Raussendorf}}\ and\ \bibinfo {author} {\bibfnamefont {H.~J.}\ \bibnamefont {Briegel}},\ }\href@noop {} {\bibfield  {journal} {\bibinfo  {journal} {Physical review letters}\ }\textbf {\bibinfo {volume} {86}},\ \bibinfo {pages} {5188} (\bibinfo {year} {2001})}\BibitemShut {NoStop}%
\bibitem [{\citenamefont {Bennett}\ \emph {et~al.}(1993)\citenamefont {Bennett}, \citenamefont {Brassard}, \citenamefont {Cr\'epeau}, \citenamefont {Jozsa}, \citenamefont {Peres},\ and\ \citenamefont {Wootters}}]{bennett1993teleporting}%
  \BibitemOpen
  \bibfield  {author} {\bibinfo {author} {\bibfnamefont {C.~H.}\ \bibnamefont {Bennett}}, \bibinfo {author} {\bibfnamefont {G.}~\bibnamefont {Brassard}}, \bibinfo {author} {\bibfnamefont {C.}~\bibnamefont {Cr\'epeau}}, \bibinfo {author} {\bibfnamefont {R.}~\bibnamefont {Jozsa}}, \bibinfo {author} {\bibfnamefont {A.}~\bibnamefont {Peres}}, \ and\ \bibinfo {author} {\bibfnamefont {W.~K.}\ \bibnamefont {Wootters}},\ }\href {\doibase 10.1103/PhysRevLett.70.1895} {\bibfield  {journal} {\bibinfo  {journal} {Phys. Rev. Lett.}\ }\textbf {\bibinfo {volume} {70}},\ \bibinfo {pages} {1895} (\bibinfo {year} {1993})}\BibitemShut {NoStop}%
\bibitem [{\citenamefont {Janotta}\ and\ \citenamefont {Hinrichsen}(2014)}]{MR3241077}%
  \BibitemOpen
  \bibfield  {author} {\bibinfo {author} {\bibfnamefont {P.}~\bibnamefont {Janotta}}\ and\ \bibinfo {author} {\bibfnamefont {H.}~\bibnamefont {Hinrichsen}},\ }\href {\doibase 10.1088/1751-8113/47/32/323001} {\bibfield  {journal} {\bibinfo  {journal} {J. Phys. A}\ }\textbf {\bibinfo {volume} {47}},\ \bibinfo {pages} {323001, 32} (\bibinfo {year} {2014})}\BibitemShut {NoStop}%
\bibitem [{\citenamefont {Sorkin}(1994)}]{sorkin1994quantum}%
  \BibitemOpen
  \bibfield  {author} {\bibinfo {author} {\bibfnamefont {R.~D.}\ \bibnamefont {Sorkin}},\ }\href@noop {} {\bibfield  {journal} {\bibinfo  {journal} {Modern Physics Letters A}\ }\textbf {\bibinfo {volume} {9}},\ \bibinfo {pages} {3119} (\bibinfo {year} {1994})}\BibitemShut {NoStop}%
\bibitem [{\citenamefont {Lee}\ and\ \citenamefont {Barrett}(2015)}]{lee2015computation}%
  \BibitemOpen
  \bibfield  {author} {\bibinfo {author} {\bibfnamefont {C.~M.}\ \bibnamefont {Lee}}\ and\ \bibinfo {author} {\bibfnamefont {J.}~\bibnamefont {Barrett}},\ }\href@noop {} {\bibfield  {journal} {\bibinfo  {journal} {New Journal of Physics}\ }\textbf {\bibinfo {volume} {17}},\ \bibinfo {pages} {083001} (\bibinfo {year} {2015})}\BibitemShut {NoStop}%
\bibitem [{\citenamefont {Lee}\ and\ \citenamefont {Selby}(2016)}]{lee2016deriving}%
  \BibitemOpen
  \bibfield  {author} {\bibinfo {author} {\bibfnamefont {C.~M.}\ \bibnamefont {Lee}}\ and\ \bibinfo {author} {\bibfnamefont {J.~H.}\ \bibnamefont {Selby}},\ }\href@noop {} {\bibfield  {journal} {\bibinfo  {journal} {New Journal of Physics}\ }\textbf {\bibinfo {volume} {18}},\ \bibinfo {pages} {093047} (\bibinfo {year} {2016})}\BibitemShut {NoStop}%
\bibitem [{\citenamefont {Lee}\ and\ \citenamefont {Hoban}(2016)}]{lee2016bounds}%
  \BibitemOpen
  \bibfield  {author} {\bibinfo {author} {\bibfnamefont {C.~M.}\ \bibnamefont {Lee}}\ and\ \bibinfo {author} {\bibfnamefont {M.~J.}\ \bibnamefont {Hoban}},\ }\href@noop {} {\bibfield  {journal} {\bibinfo  {journal} {Proceedings of the Royal Society A: Mathematical, Physical and Engineering Sciences}\ }\textbf {\bibinfo {volume} {472}},\ \bibinfo {pages} {20160076} (\bibinfo {year} {2016})}\BibitemShut {NoStop}%
\bibitem [{\citenamefont {D'Ariano}\ \emph {et~al.}(2017)\citenamefont {D'Ariano}, \citenamefont {Chiribella},\ and\ \citenamefont {Perinotti}}]{d2017quantum}%
  \BibitemOpen
  \bibfield  {author} {\bibinfo {author} {\bibfnamefont {G.~M.}\ \bibnamefont {D'Ariano}}, \bibinfo {author} {\bibfnamefont {G.}~\bibnamefont {Chiribella}}, \ and\ \bibinfo {author} {\bibfnamefont {P.}~\bibnamefont {Perinotti}},\ }\href@noop {} {\emph {\bibinfo {title} {Quantum theory from first principles: an informational approach}}}\ (\bibinfo  {publisher} {Cambridge University Press},\ \bibinfo {year} {2017})\BibitemShut {NoStop}%
\bibitem [{\citenamefont {Chiribella}\ \emph {et~al.}(2010)\citenamefont {Chiribella}, \citenamefont {D’Ariano},\ and\ \citenamefont {Perinotti}}]{chiribella2010probabilistic}%
  \BibitemOpen
  \bibfield  {author} {\bibinfo {author} {\bibfnamefont {G.}~\bibnamefont {Chiribella}}, \bibinfo {author} {\bibfnamefont {G.~M.}\ \bibnamefont {D’Ariano}}, \ and\ \bibinfo {author} {\bibfnamefont {P.}~\bibnamefont {Perinotti}},\ }\href@noop {} {\bibfield  {journal} {\bibinfo  {journal} {Physical Review A}\ }\textbf {\bibinfo {volume} {81}},\ \bibinfo {pages} {062348} (\bibinfo {year} {2010})}\BibitemShut {NoStop}%
\bibitem [{\citenamefont {Chiribella}\ \emph {et~al.}(2011)\citenamefont {Chiribella}, \citenamefont {D'Ariano},\ and\ \citenamefont {Perinotti}}]{chiribella2011informational}%
  \BibitemOpen
  \bibfield  {author} {\bibinfo {author} {\bibfnamefont {G.}~\bibnamefont {Chiribella}}, \bibinfo {author} {\bibfnamefont {G.~M.}\ \bibnamefont {D'Ariano}}, \ and\ \bibinfo {author} {\bibfnamefont {P.}~\bibnamefont {Perinotti}},\ }\href {\doibase 10.1103/PhysRevA.84.012311} {\bibfield  {journal} {\bibinfo  {journal} {Phys. Rev. A}\ }\textbf {\bibinfo {volume} {84}},\ \bibinfo {pages} {012311} (\bibinfo {year} {2011})}\BibitemShut {NoStop}%
\bibitem [{\citenamefont {Abrams}\ and\ \citenamefont {Lloyd}(1998)}]{abrams1998nonlinear}%
  \BibitemOpen
  \bibfield  {author} {\bibinfo {author} {\bibfnamefont {D.~S.}\ \bibnamefont {Abrams}}\ and\ \bibinfo {author} {\bibfnamefont {S.}~\bibnamefont {Lloyd}},\ }\href@noop {} {\bibfield  {journal} {\bibinfo  {journal} {Physical Review Letters}\ }\textbf {\bibinfo {volume} {81}},\ \bibinfo {pages} {3992} (\bibinfo {year} {1998})}\BibitemShut {NoStop}%
\bibitem [{\citenamefont {Aaronson}(2004)}]{aaronson2004quantum}%
  \BibitemOpen
  \bibfield  {author} {\bibinfo {author} {\bibfnamefont {S.}~\bibnamefont {Aaronson}},\ }\href@noop {} {\bibfield  {journal} {\bibinfo  {journal} {arXiv preprint quant-ph/0401062}\ } (\bibinfo {year} {2004})}\BibitemShut {NoStop}%
\bibitem [{\citenamefont {Aaronson}(2005)}]{aaronson2005quantum}%
  \BibitemOpen
  \bibfield  {author} {\bibinfo {author} {\bibfnamefont {S.}~\bibnamefont {Aaronson}},\ }\href@noop {} {\bibfield  {journal} {\bibinfo  {journal} {Proceedings of the Royal Society A: Mathematical, Physical and Engineering Sciences}\ }\textbf {\bibinfo {volume} {461}},\ \bibinfo {pages} {3473} (\bibinfo {year} {2005})}\BibitemShut {NoStop}%
\bibitem [{\citenamefont {Baumann}\ \emph {et~al.}(2021)\citenamefont {Baumann}, \citenamefont {Del~Santo}, \citenamefont {Smith}, \citenamefont {Giacomini}, \citenamefont {Castro-Ruiz},\ and\ \citenamefont {Brukner}}]{baumann2021generalized}%
  \BibitemOpen
  \bibfield  {author} {\bibinfo {author} {\bibfnamefont {V.}~\bibnamefont {Baumann}}, \bibinfo {author} {\bibfnamefont {F.}~\bibnamefont {Del~Santo}}, \bibinfo {author} {\bibfnamefont {A.~R.}\ \bibnamefont {Smith}}, \bibinfo {author} {\bibfnamefont {F.}~\bibnamefont {Giacomini}}, \bibinfo {author} {\bibfnamefont {E.}~\bibnamefont {Castro-Ruiz}}, \ and\ \bibinfo {author} {\bibfnamefont {C.}~\bibnamefont {Brukner}},\ }\href@noop {} {\bibfield  {journal} {\bibinfo  {journal} {Quantum}\ }\textbf {\bibinfo {volume} {5}},\ \bibinfo {pages} {524} (\bibinfo {year} {2021})}\BibitemShut {NoStop}%
\bibitem [{\citenamefont {Srikanth}(2022)}]{srikanth2022operational}%
  \BibitemOpen
  \bibfield  {author} {\bibinfo {author} {\bibfnamefont {R.}~\bibnamefont {Srikanth}},\ }\href {\doibase 10.1103/PhysRevA.106.012221} {\bibfield  {journal} {\bibinfo  {journal} {Phys. Rev. A}\ }\textbf {\bibinfo {volume} {106}},\ \bibinfo {pages} {012221} (\bibinfo {year} {2022})}\BibitemShut {NoStop}%
\bibitem [{\citenamefont {Barrett}\ \emph {et~al.}(2019)\citenamefont {Barrett}, \citenamefont {de~Beaudrap}, \citenamefont {Hoban},\ and\ \citenamefont {Lee}}]{barrett2019computational}%
  \BibitemOpen
  \bibfield  {author} {\bibinfo {author} {\bibfnamefont {J.}~\bibnamefont {Barrett}}, \bibinfo {author} {\bibfnamefont {N.}~\bibnamefont {de~Beaudrap}}, \bibinfo {author} {\bibfnamefont {M.~J.}\ \bibnamefont {Hoban}}, \ and\ \bibinfo {author} {\bibfnamefont {C.~M.}\ \bibnamefont {Lee}},\ }\href@noop {} {\bibfield  {journal} {\bibinfo  {journal} {npj Quantum Information}\ }\textbf {\bibinfo {volume} {5}},\ \bibinfo {pages} {1} (\bibinfo {year} {2019})}\BibitemShut {NoStop}%
\bibitem [{\citenamefont {Kumar}\ \emph {et~al.}(2025)\citenamefont {Kumar}, \citenamefont {Singh},\ and\ \citenamefont {Srikanth}}]{kumar2025on}%
  \BibitemOpen
  \bibfield  {author} {\bibinfo {author} {\bibfnamefont {V.}~\bibnamefont {Kumar}}, \bibinfo {author} {\bibfnamefont {M.~P.}\ \bibnamefont {Singh}}, \ and\ \bibinfo {author} {\bibfnamefont {R.}~\bibnamefont {Srikanth}},\ }\href {\doibase 10.12743/quanta.91} {\bibfield  {journal} {\bibinfo  {journal} {Quanta}\ }\textbf {\bibinfo {volume} {14}},\ \bibinfo {pages} {38} (\bibinfo {year} {2025})}\BibitemShut {NoStop}%
\bibitem [{\citenamefont {Gisin}(1990)}]{gisin1990weinberg}%
  \BibitemOpen
  \bibfield  {author} {\bibinfo {author} {\bibfnamefont {N.}~\bibnamefont {Gisin}},\ }\href@noop {} {\bibfield  {journal} {\bibinfo  {journal} {Physics Letters A}\ }\textbf {\bibinfo {volume} {143}},\ \bibinfo {pages} {1} (\bibinfo {year} {1990})}\BibitemShut {NoStop}%
\bibitem [{\citenamefont {Polchinski}(1991)}]{polchinski1991weinberg}%
  \BibitemOpen
  \bibfield  {author} {\bibinfo {author} {\bibfnamefont {J.}~\bibnamefont {Polchinski}},\ }\href@noop {} {\bibfield  {journal} {\bibinfo  {journal} {Physical Review Letters}\ }\textbf {\bibinfo {volume} {66}},\ \bibinfo {pages} {397} (\bibinfo {year} {1991})}\BibitemShut {NoStop}%
\end{thebibliography}

    %merlin.mbs apsrev4-1.bst 2010-07-25 4.21a (PWD, AO, DPC) hacked
%Control: key (0)
%Control: author (8) initials jnrlst
%Control: editor formatted (1) identically to author
%Control: production of article title (-1) disabled
%Control: page (0) single
%Control: year (1) truncated
%Control: production of eprint (0) enabled
\providecommand{\noopsort}[1]{}\providecommand{\singleletter}[1]{#1}%

\appendix
\section{DAGs and causal structure \label{sec:DAG}}
A DAG graphically encodes the conditional independences among a given set of random variables. For any probability distribution that is compatible with a given causal structure, represented by DAG $\mathfrak{G}$, the causal Markov condition, a core principle of classical causal structure, stipulates thus: $P(T | \textrm{Nd}(T), \textrm{Pa}(T)) = P(T | \textrm{Pa}(T))$, i.e., conditioned on its parents $\textrm{Pa}(T)$, the probability of a given variable $T$ is independent of its non-descendents $\textrm{Nd}(T)$. This can be represented by the notation $(T \independent~\textrm{Nd}(T) | \textrm{Pa}(T))$. Extending this to $n$ vertices $T_1, T_2, \cdots, T_n$ of graph $G$, the causal Markov condition is equivalent to requiring that the joint probability distribution factorizes as
\begin{equation}
P(X_1, X_2, \cdots, X_n) = \Pi_j P(X_j | \textrm{Pa}(X_j)).
\label{eq:ShriMarudurAmmaMarkov}
\end{equation}
This generalizes the Reichenbach principle of common cause. The conditional independences implied by the causal Markov condition can be read off with relative ease using the following graphical criterion, called d-separation \cite{wood2015lesson}.

\section{Causal structure in Weinberg's nonlinear theory \label{sec:weinberg}} 
It is instructive to consider the application of these results to a nonlinear quantum theory proposed by Weinberg \cite{weinberg1989testing}. It is known to lead to physically implausible effects. In particular, Ref. \cite{abrams1998nonlinear} exploits this nonlinearity to design an algorithm to efficiently solve $\mathbf{NP}$-complete and \#$\mathbf{P}$ problems. Earlier, Weinberg's nonlinear theory was shown to lead to superluminal signaling \cite{gisin1990weinberg, polchinski1991weinberg}. This comes about essentially because the theory allows the existence of a non-bilinear Hamiltonian, which evolves both states $\ket{0}$ and $\ket{1}$ to the state $\ket{1}$. 

Suppose Alice and Bob start by sharing a Bell state $\ket{\Phi^{+}} \equiv \frac{1}{\sqrt{2}}(\ket{0,0}+\ket{1,1})_{AB}$. Alice either does nothing, or applies the above nonlinear transformation. The latter action implements the transformation 
$\frac{1}{\sqrt{2}}(\ket{0,0}+\ket{1,1})_{AB} \longrightarrow \frac{1}{\sqrt{2}}(\ket{1,0}+\ket{1,1})_{AB}$. Thus, depending on Alice's choice, Bob's particle is left in the state $\frac{\mathbb{I}}{2}$ or $\ket{+}\equiv \frac{1}{\sqrt{2}}(\ket{0}+\ket{1})$, which would allow Alice to signal to Bob.

The objective signal that Bob receives can be explained by a classical causal model,  such as the DAG of Fig. \ref{fig:Arrow} that includes superluminal arrows. The signal cannot be explained by quantum causal model, which corresponds to a DAG such as Fig. \ref{fig:bell} that conforms to RC. By Proposition \ref{thm:onto}, a subjective interpretation of the nonlinear QM is ruled out. This provides an instance of a nonclassical theory where the system's state inherently favors one type of interpretation-- the realist, in this case. Evidently, this unambiguous interpretation of the quantum state comes for the price of superluminal signaling. \color{black}

\end{document}